\documentclass[journal]{IEEEtran}
\usepackage{tensor}
\usepackage{graphicx}
\usepackage{tabularx}
\usepackage[cmex10]{amsmath}
\usepackage{amsthm}
\usepackage{amssymb}
\usepackage{bm}
\usepackage{algorithm}
\usepackage{algorithmic}
\usepackage{setspace}
\usepackage{stfloats}
\usepackage{enumerate}
\usepackage{cases}
\usepackage{multirow}
\usepackage{mathrsfs}
\usepackage{array}
\usepackage{color}
\usepackage{verbatim}
\usepackage{extpfeil} 
\usepackage{booktabs}
\usepackage{graphicx}
\usepackage{tabularx}
\usepackage{epstopdf}
\usepackage{textcomp}
\usepackage{epsfig,epsf,color,balance,cite}
\usepackage{amsmath}
\usepackage{epsfig,epsf,color,balance,cite}
\usepackage{textcomp}
\usepackage{url}
\usepackage{tikz}  
\usepackage[colorlinks,citecolor=blue,linkcolor=blue,urlcolor=blue]{hyperref}

\usepackage{booktabs,tabularx,makecell}

\renewcommand{\arraystretch}{1.05}

\def\BibTeX{{\rm B\kern-.05em{\sc i\kern-.025em b}\kern-.08em
		T\kern-.1667em\lower.7ex\hbox{E}\kern-.125emX}}
\begin{document}

\makeatletter
\newcommand{\rmnum}[1]{\romannumeral #1}
\newcommand{\Rmnum}[1]{\expandafter \@slowromancap \romannumeral #1@}
\makeatother

\title{Reimagining Wireless Connectivity:\\The FAS-RIS Synergy for 6G Smart Cities}

\author{Tuo Wu, 
            Kai-Kit Wong, \emph{Fellow, IEEE},
            Jie Tang, 
            Junteng Yao, 
            Baiyang Liu, \\ 
            Kin-Fai Tong, \emph{Fellow, IEEE}, 
            Chan-Byoung Chae, \emph{Fellow, IEEE}, 
            Matthew  C. Valenti, \emph{Fellow}, \emph{IEEE}, \\ 
            and Kwai-Man Luk, \emph{Life Fellow, IEEE}
\vspace{-7mm}

\thanks{(\textit{Corresponding authors: K.-K.Wong, and K.-F. Tong.})}
\thanks{T. Wu and K.-M. Luk are with the State Key Laboratory of Terahertz and Millimeter Waves, City University of Hong Kong, Hong Kong, China (E-mail: $\rm \{tuowu2, eekmluk\}@cityu.edu.hk$). K.-K. Wong is with the Department of Electronic and Electrical Engineering, University College London, WC1E 6BT London, U.K., and also with the Yonsei Frontier Laboratory and the School of Integrated Technology, Yonsei University, Seoul 03722, South Korea (E-mail: $\rm  kai\text{-}kit.wong@ucl.ac.uk$). J. Tang is with the School of Electronic and Information Engineering, South China University of Technology, Guangzhou 510640, China (E-mail: $\rm eejtang@scut.edu.cn$). J. Yao is with the faculty of Electrical Engineering and Computer Science, Ningbo University, Ningbo 315211, China (E-mail: $\rm yaojunteng@nbu.edu.cn$). B. Liu and  K. F. Tong are with the School of Science and Technology, Hong Kong Metropolitan University, Hong Kong SAR, China. (E-mail: $\rm \{byliu,ktong\}@hkmu.edu.hk$). C.-B. Chae is with the School of Integrated Technology, Yonsei University, Seoul 03722 Korea. (E-mail: $\rm cbchae@yonsei.ac.kr$). M. C. Valenti is with the Lane Department of Computer Science and Electrical Engineering, West Virginia University, Morgantown, USA (E-mail: $\rm valenti@ieee.org$).}  
}

\markboth{}
{}
\maketitle

\begin{abstract}
Fluid antenna system (FAS) represents the concept of treating antenna as a reconfigurable physical-layer resource to broaden system design and network optimization and inspire next-generation reconfigurable antennas. FAS can unleash new degree of freedom (DoF) via antenna reconfigurations for novel spatial diversity. Reconfigurable intelligent surfaces (RISs) on the other hand can reshape wireless propagation environments but often face limitations from double path-loss and minimal signal processing capability when operating independently. This article envisions a transformative FAS-RIS integrated architecture for future smart city networks, uniting the adaptability of FAS with the environmental control of RIS. The proposed framework has five key applications: FAS-enabled base stations (BSs) for large-scale beamforming, FAS-equipped user devices with finest spatial diversity, and three novel RIS paradigms---fluid RIS (FRIS) with reconfigurable elements, FAS-embedded RIS as active relays, and enormous FAS (E-FAS) exploiting surface waves on facades to re-establish line-of-sight (LoS) communication. A two-timescale control mechanism coordinates network-level beamforming with rapid, device-level adaptation. Applications spanning from simultaneous wireless information and power transfer (SWIPT) to integrated sensing and communications (ISAC), with challenges in co-design, channel modeling, and optimization, are discussed. This article concludes with simulation results demonstrating the robustness and effectiveness of the FAS-RIS system.
\end{abstract}


\section{Introduction} 
\IEEEPARstart{S}{ixth}-generation (6G) wireless networks envision a transformative leap beyond fifth-generation (5G), demanding hyper-reliable low-latency communications (HRLLC), integrated sensing and communications (ISAC) for simultaneous environmental perception and data transmission, ubiquitous communication, edge artificial intelligence (AI), and massive internet-of-things (IoT) connectivity in smart cities, industrial and unpredictable environments \cite{IMT-2030-white}. Together with immersive communications, they form the six key use cases for 6G.

Innovations are required to push towards the 6G frontiers and over the past few years, reconfigurable intelligent surfaces (RISs) \cite{HuangC19} have become one of the hottest topics in wireless communications. Consisted of many phase-tunable elements, RIS possesses the ability to redirect incident radio waves for smart reflection, leading to improved reliability, enhanced rate and extended coverage. Practical RISs, however, suffer from serious drawbacks. For example, the number of elements is highly restricted in implementation and besides, they normally come with finite-resolution phase control (e.g., $1$-$2$ bits). The supposed gains of RIS are therefore rarely realizable.

Thinking beyond passive engineering of wireless environments, recent interest in utilizing reconfigurable antennas for wireless communications appears to be a timely effort \cite{Heath25}. As a matter of fact, in 2020 \cite{FAS21}, Wong {\em et al.}~proposed the concept of Fluid Antenna Systems (FASs) which treats antennas as a reconfigurable physical-layer resource (an engineering object) to broaden system design and network optimization, inspiring next-generation reconfigurable antennas. FAS may come in a variety of forms such as liquid metal, mechanical actuators, electronically switchable pixels, and programmable metasurfaces \cite{Tong-2025}, and is thus hardware-agnostic. Recent articles discussing the potential for FAS can be found in \cite{Lu-2025}.

Emphasizing on the features of shape and position flexibility in antennas, FAS empowers the physical layer with greater spatial degrees of freedom (DoFs), leading to new modulation, multiple access and signal processing schemes for tremendous performance gains \cite{New24}. Motivated by this, FAS can be the key to make RIS a practically attractive solution that could finally reduce the burden of base stations (BSs) and lessen the need for network densification when going up the frequency band. 

This article discusses the FAS-RIS architecture that advocates a synergistic integration of FAS and RIS, moving beyond simple co-existence  \cite{JYao} to form a multi-layered, intelligent wireless fabric. We propose three paradigms for FAS-RIS:
\begin{itemize}
\item Fluid RIS (FRIS) \cite{Xiao25}---This type integrates `fluid' reconfigurability into each element of RIS.
\item FAS-embedded RIS---This type embeds `fluid' reconfigurability into each element of an intelligent surface but instead of serving as a passive reflector, the RIS acts as an active, intelligent relay processor with each element capable of reconfiguring its position for reception and choosing another position for signal forwarding.
\item Enormous FAS (E-FAS), a.k.a.~surface-wave FAS (SW-FAS) \cite{LiuQ24,Wong-2025efas}---This paradigm leverages surface waves to turn entire building facades/walls into low-loss line-of-sight (LoS) communication and sensing apertures.
\end{itemize}
	
This architecture is distinctive from the conventional simple co-existence as it necessitates a two-timescale control scheme that orchestrates network-level macro-beamforming and RIS configuration with device-level FAS adaptation. This synergy creates exceptionally resilient and efficient links, paving the way for advanced applications in complex smart-city scenarios, including road safety through reliable autonomous vehicle connectivity, reducing carbon emissions through energy-efficient building-integrated communication, and etc.

This article aims to present a systematic overview of the FAS-RIS framework. We first briefly review its possible hardware architectures and their trade-offs. Then we discuss the synergy of FAS and RIS in the context of smart cities and unpack the three FAS-RIS paradigms in Section \ref{sec:synergy}. Section \ref{sec:open} identifies simultaneous wireless information and power transfer (SWIPT) \cite{LZhou} and integrated sensing and communication (ISAC) \cite{CWang24} as some key smart city applications for FAS-RIS and then dives into the open challenges. Section \ref{sec:case} provides simulation results illustrating the robustness and efficiency of FAS-RIS systems across diverse environments. In Section \ref{sec:conclude}, we conclude and call for further research on this topic.

\begin{figure*}[]  
\centering
\includegraphics[width=0.99\textwidth]{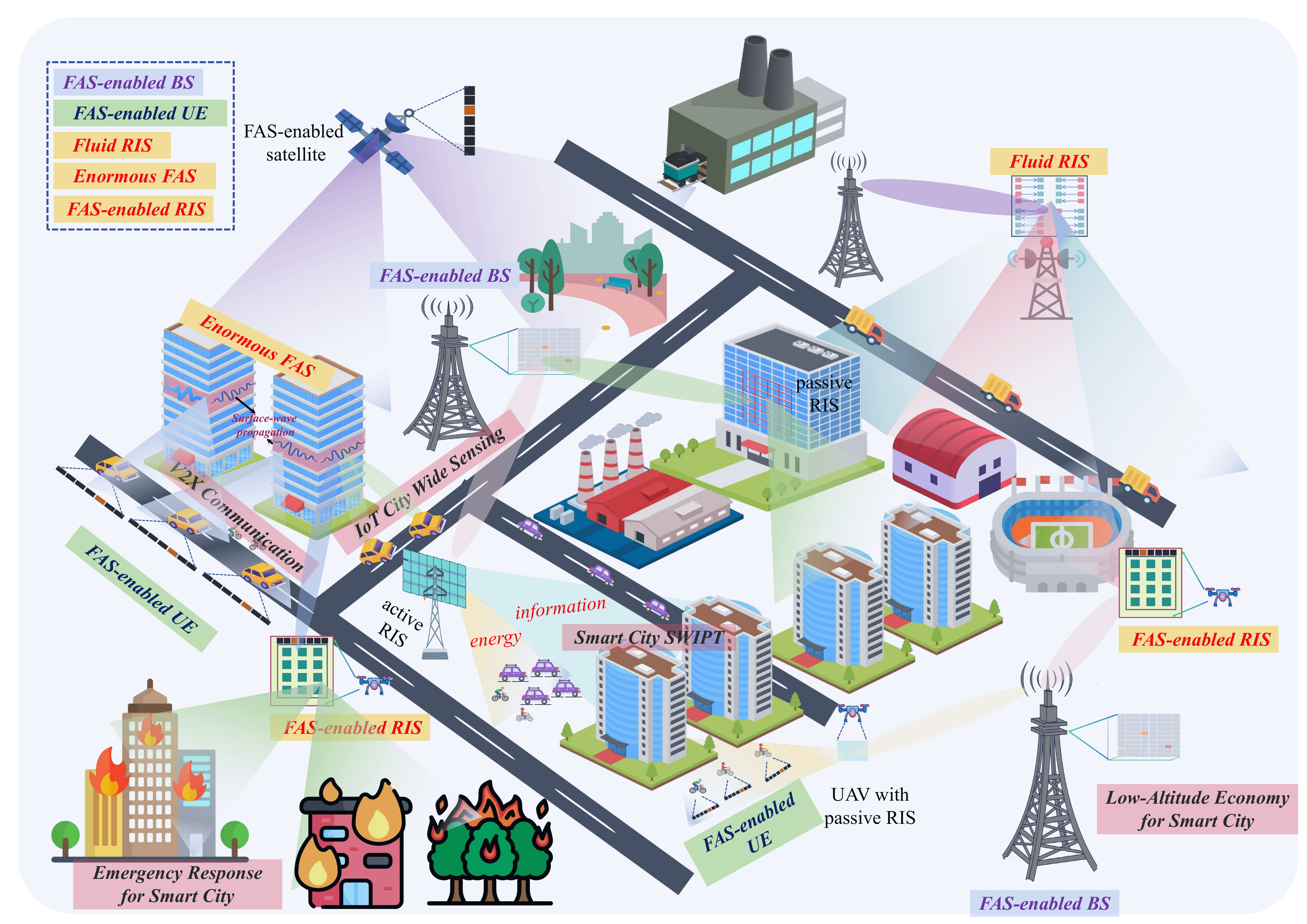}
\caption{A comprehensive FAS-RIS system architecture for 6G smart city wireless networks, showcasing various deployment scenarios including FAS-enabled satellites, BSs, intelligent surfaces, unmanned aerial vehicles (UAVs), and mobile user equipments (UEs).}\label{fig1}
\vspace{-2mm}
\end{figure*}

\section{FAS-RIS Architectures}\label{sec:synergy}
Fig.~\ref{fig1} depicts a comprehensive 6G wireless network ecosystem that integrates several FAS-RIS configurations across diverse deployment scenarios, including terrestrial BSs, satellite communications, UAVs, and mobile UEs. In addition, Table~\ref{table:fas_ris_overview} contrasts FAS with fixed-position antennas (FPA) at the BS/UE and compares FAS-enabled RIS with the conventional passive RIS, while also linking these design choices to applications and challenges which will be discussed later.  

\begin{table*}[]
\caption{Architectural comparison (BS/UE/RIS) with applications and challenges.}\label{table:fas_ris_overview}
\centering
\footnotesize
\renewcommand{\arraystretch}{1.05}
\begin{tabularx}{.95\textwidth}{p{2.4cm} p{3.2cm} 
				>{\centering\arraybackslash}X 
				>{\centering\arraybackslash}X 
				>{\centering\arraybackslash}X}
			\toprule
			\multicolumn{2}{c}{\textbf{Dimension}} & 
			\textbf{FAS-BS vs FPA-BS} & 
			\textbf{FAS-UE vs FPA-UE} & 
			\textbf{FAS-Enabled RIS vs Passive RIS} \\
			\midrule
			\multirow{3}{*}{\textbf{Capabilities}} 
			& Beamforming/DoF & Higher, reconfigurable & Higher, mobility-robust & Environment shaping/relay/surface-wave \\
			\cmidrule(lr){2-5}
			& Port/Position Diversity & \checkmark & \checkmark & \checkmark (structural) \\
			\cmidrule(lr){2-5}
			& Complexity/Power & Higher than FPA & Slightly higher & Varies (passive/active/E-FAS) \\
			\midrule
			\multirow{3}{*}{\textbf{Applications}} 
			& SWIPT & FAS-BS directs information + energy beams via port/position/shape reconfiguration; targets active or passive RIS 
			& UE FAS port/position/shape diversity finds energy + information sweet-spot; boosts energy harvesting (EH) and signal-to-noise ratio (SNR) 
			& Active FRIS amplifies/splits; E-FAS on-demand launchers deliver to targets \\
			\cmidrule(lr){2-5}
			& ISAC & BS crafts sensing + communication dual-purpose beams; schedules tasks (two-timescale) 
			& UE FAS fine sensing via fast port switching; robust under motion/occlusion 
			& FRIS shapes echoes; E-FAS surface-wave enables around-the-corner sensing and imaging \\
			\cmidrule(lr){2-5}
			& Satellite & On-orbit M-FAS forms agile spot beams; fast reconfiguartion; counters atmosphere/attitude; hotspot tracking 
			& UE FAS port/position/shape diversity aligns satellite beams; mitigates blockage/motion; robust links 
			& Ground FRIS co-shapes uplink (UL)/downlink (DL): UL pre-compensation, DL coherent combining and obstacle-bypass; extends coverage \\
			\midrule
			\multirow{5}{*}{\textbf{Challenges}} 
			& Architechure and Control Co-Design & Two-timescale network-level beam scheduling and coverage orchestration 
			& Fast device-level adaptation and port selection under mobility 
			& Policies for configuration/update of reflection, relaying, and surface-wave launchers; E-FAS device/panel engineering \\
			\cmidrule(lr){2-5}
			& Unified Channel Modeling & Unified models beyond field-response and correlation-based models; capture hybrid guided-radiated propagation 
			& Time-varying port correlation and local scattering statistics 
			& Cascaded channels with surface-to-space transitions \\
			\cmidrule(lr){2-5}
			& Hierarchical Channel State Information (CSI)/Training & Two-timescale estimation/feedback protocols 
			& Low-overhead local CSI acquisition and tracking 
			& Global CSI and structural-state inference for FRIS/E-FAS \\
			\cmidrule(lr){2-5}
			& Joint Beamforming and Resource & Cross-layer optimization of BS beams, FRIS states, and UE ports 
			& Port selection with quality-of-service (QoS)/latency constraints 
			& Selection of reflection/relaying/launcher activation points \\
			\cmidrule(lr){2-5}
			& Sensing and Localization & Task scheduling and sensing beam design 
			& Cooperative UE sensing and data fusion 
			& E-FAS surface-wave enabled non-LoS (NLoS) or around-the-corner imaging \\
			\bottomrule
\end{tabularx}
\end{table*}

As shown in Fig.~\ref{fig1}, this architecture underpins several key smart-city applications. Autonomous vehicles leverage FAS-enabled roadside units for reliable vehicle-to-everything (V2X) connectivity even in dense urban canyons; UAV equipped with FRIS can be deployed for emergency response; E-FAS-coated building facades provide energy-efficient communication infrastructure; also, FRIS enables IoT sensor networks to harvest energy wirelessly, alleviating the need for battery replacement; and the ISAC capabilities of FAS-RIS support intelligent traffic management and public-safety monitoring.


\subsection{Implementation Technologies}
Successful integration of FAS and RIS hinges on the availability of diverse FAS implementation technologies, see \cite{Tong-2025,Lu-2025}, \cite[Section VI]{New24}, each acting as a building block for different components of the overall architecture, as summarized in Table~\ref{table:fas_implementations}. Liquid-metal-based FAS, which repositions conductive fluid through microfluidic channels, provides robust reconfigurability with continuous states and is very well suited to mobile UEs. Mechanically movable FAS, which employs actuators to physically reposition antenna elements, offers large-scale reconfigurability but with very slow speeds, making it particularly suitable for static deployments such as building-mounted RIS. Reconfigurable pixel-based FAS uses electronic switches (e.g., PIN diodes) to toggle between discrete antenna states on a nanosecond timescale, making it well suited to power- and space-constrained mobile UEs and BSs. Meta-fluid antenna (M-FAS) leverages electronically controlled meta-atom arrays to emulate fluid-like electromagnetic behavior with nanosecond-scale reconfiguration and very high spatial DoF, making it suitable for a wide range of systems. Then E-FAS, utilizing surface-wave launchers with nanosecond-scale switching and very high spatial DoF, is particularly promising for large-scale deployments such as LED screens.

Evidently, each implementation exhibits distinct trade-offs in reconfiguration speed, hardware complexity, power consumption, bandwidth, and achievable performance, enabling designers to tailor FAS-RIS deployments to the requirements of specific system components and application scenarios.

\begin{table}[t]
\caption{Comparison of implementation technologies.}\label{table:fas_implementations}
\centering
\scriptsize
\renewcommand{\arraystretch}{1.12}
\setlength{\tabcolsep}{3.2pt} 
\begin{tabularx}{1\columnwidth}{%
>{\raggedright\arraybackslash}p{2.55cm}%
*{5}{>{\centering\arraybackslash}X}}
\toprule
\textbf{\makecell[l]{Characteristics}} & \textbf{Liquid Metal} & \textbf{Mechanical} & \textbf{Pixel-based} & \textbf{M-FAS} & \textbf{E-FAS}\\
\midrule
\textbf{\makecell[l]{Reconfig. Speed}} & Slow (s) & Very slow (s) & Fast (ns) & Fast (ns) & Fast (ns)\\
\cmidrule(lr){1-6}
			\textbf{\makecell{Control Method}}
			& Micro  pumps
			& Motors
			& Electronic switches
			& Meta-atom control
			& Surface-wave launchers\\
\cmidrule(lr){1-6}
			\textbf{\makecell{Precision}}
			& Medium
			& Low
			& High
			& High
			& High\\
			\midrule

			\textbf{\makecell[l]{Power Consumption}}
			& Medium & High & Low & Low & Medium\\
			\cmidrule(lr){1-6}
			\textbf{\makecell[l]{Hardware Complexity}}
			& High   & High & Medium & Medium & Medium\\
			\midrule
			\textbf{\textbf{\makecell[l]{Reconfig. States}}}
			& Continuous & Continuous & Discrete (few states) & Discrete (more states) & Continuous\\
			\cmidrule(lr){1-6}
			\textbf{\makecell[l]{Spatial DoF}}
			& Medium & Medium & Limited & Very high & Very high\\
			\cmidrule(lr){1-6}
			\textbf{\makecell[l]{Bandwidth}}
			& 22-39 GHz & Wide  & 6-7 GHz & 26-27 GHz & 34-86 GHz\\
			\cmidrule(lr){1-6}
			\textbf{\makecell[l]{Terminal}}
			& UE & BS, RIS & Universal & Universal & Indoor \& outdoor RIS, e.g., LED screens\\
			\bottomrule
\end{tabularx}
\end{table}

\subsection{FAS-enabled BS}
As depicted in Fig.~\ref{fig1}, deploying FAS at the BS significantly enhances the beamforming gain of FAS-RIS systems through dynamic reconfiguration. For BS deployment, mechanically movable FAS and M-FAS are particularly attractive, as they can exploit the relatively generous space and power budget at the BS to realize large-scale physical reconfiguration and improved performance though movable ones could only react to long-term channel statistics. Pixel-based FAS can be effective too in this case but its optimization would be more complex. Compared with the conventional FPA counterpart, the FAS-BS setup in Table~\ref{table:fas_ris_overview} provides real-time port/position diversity and reconfigurability, enabling flexible concentration and steering of signal power. This capability helps to compensate for the double-path loss in cascaded channels of RIS-aided environments, improve signal coverage and strength, and maintain a high SNR even in complex propagation environments, thereby enhancing the quality and stability of communication links.

The FAS at the BS, together with the RIS elements it coordinates, operates under a two-layer control architecture: the FAS performs adaptive transmit beamforming, while passive or active RIS deployments along dominant paths (e.g., building facades, roadsides, intersections, UAV platforms) shape the reflected field. As usual, passive RIS reconstructs NLoS links and can suppress interference while active RIS enables joint information and energy reflection for SWIPT. Moreover, UAV-mounted RIS can extend angular coverage. By co-scheduling FAS beam updates and RIS phase profiles within a common control loop---assigning coarse, wide-area steering to the FAS (mechanical, pixel or M-FAS) and fine, last-meter redirection to the RIS---the system can achieve higher end-to-end SNR, more uniform coverage, and mobility-robust links with modest training overhead, as summarized in Table~\ref{table:fas_ris_overview}.

 \subsection{FAS-enabled UE} 
In the FAS-RIS architecture, the FAS-enabled UE serves as the crucial final stage, whose primary role is to extract fine-grained spatial diversity at the receiver. While the BS and RIS shape the macro-scale propagation environment, the UE's fluid antenna complements this by rapidly switching among many closely spaced physical positions to locate and operate from the point with the best instantaneous signal quality. This ability to exploit spatial diversity within a compact form factor is the key advantage of the FAS-RIS synergy. It translates into higher average SNR, markedly lower outage probability, and inherent robustness to multipath fading and physical blockages---performance gains that a fixed-antenna UE cannot match, even in the same RIS-assisted environment. To satisfy the stringent power and speed constraints of mobile devices, compact pixel-based FAS and M-FAS implementations are preferred due to their microsecond-scale reconfiguration, whereas larger platforms such as the vehicles in Fig.~\ref{fig1} can employ liquid-metal or mechanically movable FAS to enhance both communication reliability and sensing for ISAC. In the uplink when UE is the transmitter, the same benefits remain.

To fully exploit this spatial diversity, a two-level control mechanism is adopted. At the fast timescale, the UE executes continuous port switching (the microsecond-scale ``fast loop'') to preserve the best possible link in the presence of local channel fluctuations. When larger-scale channel changes occur, such as a user turning a corner---the device notifies the network to trigger slower updates of the RIS and BS configurations, re-optimizing the macro environment. This seamless ``macro-shaping plus micro-sampling'' collaboration is reflected in the smart-city scenarios of Fig.~\ref{fig1}: a pedestrian's UE mitigates dead zones by locking onto local sweet spots created by passive RIS; it leverages active RIS for simultaneous high-rate data and SWIPT; and in vehicles, it sustains robust links under rapid motion. This intelligent, multi-level adaptation makes the FAS-enabled UE a cornerstone of the FAS-RIS vision for smart cities, with key benefits summarized in Table~\ref{table:fas_ris_overview}.
 
\subsection{FAS-enabled RIS}
Integrating FAS directly into the RIS layer transforms it from a purely reflector into a dynamic, intelligent surface that can be realized in three distinct paradigms.

\subsubsection{FRIS \cite{Xiao25}}
The RIS's reflective elements can be made reconfigurable beyond phase control. This can be implemented by using mechanical actuators or liquid-metal technologies if smart reflection from RIS is designed to respond to only long-term channel statistics. The main advantage is direct control over the coverage of the reflected beam. By optimizing the inter-element spacing and arrangement, the FRIS can either broaden the reflected beam to serve multiple UEs over a wider area, or concentrate it into a narrow, high-intensity beam to deliver maximum power to a specific target. This adaptability is particularly attractive in dynamic scenarios in which user locations and densities fluctuate, and the system must switch between wide-area broadcasting and pinpoint energy delivery. Also, the position reconfigurability of the RIS's elements can help compensate for the loss in resolution in phase control to reduce hardware cost. On the other hand, when instantaneous CSI is available, then pixel-based or M-FAS technologies are required to reconfigure the elements of FRIS.

\subsubsection{FAS-embedded RIS} 
In the second paradigm, the passive RIS is upgraded by integrating an FAS directly onto its surface. This transforms the RIS from a purely passive reflector into an intelligent, low-power active relay or transceiver. The key advantage is that the embedded FAS provides spatial diversity for the active operations: the RIS can dynamically select one optimal location on its surface to receive a signal (e.g., from the BS) and another to transmit a new, possibly amplified, signal (e.g., to a UE), thereby mitigating local fading on both uplink and downlink segments of the relay link. This paradigm is particularly well suited to mobile platforms such as the highly versatile UAV-mounted RIS in Fig.~\ref{fig1}, which can act as an agile, flying relay station to establish robust communication links for ground users. In doing so, it alleviates the double-path-loss limitation of purely passive RIS and adds a powerful layer of spatial intelligence to signal relaying.

\subsubsection{E-FAS \cite{LiuQ24,Wong-2025efas}}
The most transformative paradigm is arguably E-FAS a.k.a.~SW-FAS, which fundamentally rethinks how signals propagate. E-FAS coats surfaces such as walls with a reconfigurable dual-purpose platform, serving as both propagation medium and emitter. Rather than relying on free-space reflections, E-FAS acting as a waveguide guides signals as low-loss surface waves along the wall, effectively creating a ``communication superhighway \cite{Wong-2021swc}.'' FAS principles are then applied to realize on-demand emitters that can radiate high-gain beams from arbitrary points on the surface into free space. This yields two key benefits: a drastic reduction in propagation loss by re-establishing the LoS and a substantial mitigation of interference arising from multiple free-space reflections, forming a robust, high-capacity communication fabric. In this way, entire structures become intelligent radiating entities, representing a near-ultimate vision of a smart radio environment, as reflected in the RIS comparison in Table~\ref{table:fas_ris_overview}.

\section{Applications and Open challenges}\label{sec:open}
Next, we outline the key application opportunities and open challenges of FAS-RIS in next-generation communication systems. In Table~\ref{table:fas_ris_overview}, we consolidate the architectural comparison from Section \ref{sec:synergy} with representative applications and research challenges, providing an at-a-glance overview that links design choices to use cases and outstanding problems.

\subsection{Potential Applications}
\subsubsection{SWIPT}
The FAS-RIS architecture is well suited to enhancing SWIPT \cite{LZhou}, as shown by the active-RIS scenario in Fig.~\ref{fig1}. In this setup, the FAS-enabled BS initiates the process by dynamically adjusting its antenna ports, positions, or shape to transmit a highly focused, high-power energy beam toward a strategically placed active RIS. Unlike its passive counterpart, the active RIS can amplify and split the incident signal, forming separate, optimized beams for information and energy transfer to different users. At the receiver side, FAS-enabled UEs exploit their own spatial diversity, switching ports to lock onto the strongest peak of the incoming energy or information wavefront. This coordinated three-stage synergy---FAS-BS beamforming, active-RIS amplification and splitting, and FAS-UE reception---effectively mitigates the severe path loss and spatial constraints that limit traditional SWIPT systems, enabling robust and efficient wireless charging in smart-city environments, as summarized in Table~\ref{table:fas_ris_overview}.

\subsubsection{ISAC}
The FAS-RIS architecture can naturally enhance ISAC by creating a highly controllable sensing environment \cite{CWang24}. As shown in Fig.~\ref{fig1}, the FAS-enabled BS can generate tailored sensing beams; different types of RIS (passive, active, or fluid) can reflect and shape the echoes from targets; and FAS-enabled UEs can perform fine-grained sensing from their specific locations. This multi-point, collaborative sensing framework improves target detection and parameter estimation compared with conventional ISAC architectures.

Among the various options, the E-FAS paradigm is particularly transformative. It turns a building facade into a massive, active aperture for both communication and sensing by guiding signals as surface waves along the structure. The same surface-wave field simultaneously carries communication data and probes the environment. For example, a surface wave can maintain a robust link with a vehicle further down the street, while its interaction with another object generates backscatter that is captured along the surface for high-resolution sensing. This unique mechanism enables the network to communicate and sense ``around corners,'' greatly extending its perceptual reach beyond the BS LoS and creating a distributed sensing fabric well suited to smart-city applications.

\subsubsection{Other Applications}
Beyond SWIPT and ISAC, the FAS-RIS framework is a powerful enabler for a range of emerging applications, including spatial multiplexing and new ways for multiple access \cite[Section V]{New24}, beam steering for near-field communications, and dynamic channel shaping for physical-layer security. Besides, the same framework provides a robust approach to integrating terrestrial and non-terrestrial networks, as illustrated by the FAS-enabled satellite in Fig.~\ref{fig1}. Onboard the satellite, an FAS can dynamically reconfigure its radiation pattern to mitigate atmospheric distortions and form agile spot beams, improving coverage and link reliability without relying on complex mechanical components. On the ground, the surrounding FAS-RIS infrastructure can pre-compensate uplink signals or coherently combine downlink signals, creating a resilient, multi-layered network. This synergy is important for extending 6G services to remote and underserved regions, supporting ubiquitous connectivity for global IoT, emergency communications, and disaster response.

\subsection{Challenges and Opportunities}
Despite its notable advantages, realizing the full potential of the FAS-RIS vision will require tackling several distinctive research challenges that arise from its multi-layered and technologically diverse architecture.

\subsubsection{Architecture and Control Co-Design}
The primary architectural challenge lies in the joint design of a three-tier control system, which must coordinate the BS's macro-scale beamforming, the RIS's environment shaping (via reflection, relaying, or surface-wave propagation), and the UE's micro-scale diversity selection. A crucial open question is how to construct an efficient cross-layer control plane that operates on two distinct timescales: a slower timescale for network-level adaptation to large-scale mobility or channel statistics, and a faster timescale for device-level response to local channel fluctuations, while maintaining stability and limiting signalling overhead. In addition, deployment strategies for different FAS-RIS types---such as embedding E-FAS into building facades versus using mobile, FAS-embedded RIS on UAV platforms---must be tailored to specific urban application scenarios.

For E-FAS, this co-design challenge further extends to:
\begin{itemize}
\item[(i)] Identifying optimal surface-wave launching positions on large structures to emit radio signals to UEs;
\item[(ii)] Designing high-efficiency, wideband launchers with stable impedance and radiation characteristics;
\item[(iii)] Developing low-threshold-voltage drive electronics and bias networks to enable energy-efficient excitation;
\item[(iv)] Optimizing guided pathways with respect to propagation loss and frequency/channel partitioning;
\item[(v)] Selecting low-loss, high-permittivity (transparent) dielectric substrates for millimeter or terahertz panels; and
\item[(vi)] Engineering low-loss, phase-coherent interconnections among SW-FAS panels to support scalable deployments.
\end{itemize}

\subsubsection{Unified Channel Modeling}
Traditional channel modeling for FAS has largely relied on field-response or correlation-based models \cite{New24}, but the diverse physics and multi-layered nature of the FAS-RIS framework make these approaches increasingly inadequate. The integration of multiple FAS units with heterogeneous RIS configurations gives rise to complex cascaded channels that cannot be faithfully captured by simple geometric or correlation-based descriptions. A unified modeling framework must account for the dynamic reconfiguration of both FAS and RIS elements, the time-varying spatial correlations induced by fluid-antenna position switching, and the distinct electromagnetic characteristics of different implementations (liquid-metal, mechanical, pixel-based, M-FAS, etc.). Moreover, it must reflect the three-tier control structure, in which BS-side FAS, RIS platforms, and UE-side FAS operate on different timescales and with varying degrees of coupling. A key objective is to develop a comprehensive model that can describe the end-to-end channel across all FAS-RIS paradigms, moving beyond the limitations of current models while being tractable to support analysis and algorithm design.

\subsubsection{Hierarchical Channel Estimation}
Accurate and efficient CSI acquisition in FAS-RIS systems calls for a hierarchical, two-timescale estimation strategy \cite{HXu23}. In practice, this entails two distinct types of CSI: a fast-updating, instantaneous CSI component to support rapid device-level adaptation, and a slower-updating, statistical CSI component to guide network-level configuration. A central challenge is to design pilot and feedback mechanisms that can obtain both types of CSI with minimal signalling overhead. For E-FAS, the problem becomes even more intricate, as it requires estimating the state of a large, continuous surface and distinguishing between the surface-wave channel and the radiated free-space channel.

\subsubsection{Joint Beamforming and Resource Allocation}
Effective beamforming in FAS-RIS systems is a high-dimensional joint optimization problem spanning all three architectural layers. The core challenge is to dynamically allocate resources by jointly designing the BS beam, the RIS configuration (e.g., the reflection matrix of a passive RIS, the receive/transmit ports of an active FAS-embedded RIS, or the excitation points on an E-FAS), and the UE's port selection. This calls for new algorithms capable of solving such coupled problems in (near) real time. For E-FAS in particular, beamforming reduces to designing, on the fly, the optimal surface-wave launching pattern and excitation points to produce the desired radiation from the most effective locations on the structure.
	
\subsubsection{E-FAS-Powered Sensing and Localization}
While all the FAS-RIS paradigms can enhance localization, E-FAS offers a genuinely new opportunity to transform entire structures into high-resolution sensing platforms. The central challenge, and opportunity, is to exploit surface waves for unprecedented sensing capabilities. By analyzing how these guided surface waves interact with objects on or near the surface, E-FAS could enable continuous, high-precision tracking of vehicles and pedestrians along a building facade. Moreover, by treating the surface as a massive distributed antenna array, it can achieve ``around-the-corner'' localization and imaging of targets that lie outside the LoS of any individual BS, opening up new frontiers for smart-city sensing applications.

\section{Case Study}\label{sec:case}
\subsection{Setups}
\begin{figure*}[]  
\centering
\includegraphics[width=0.9\textwidth]{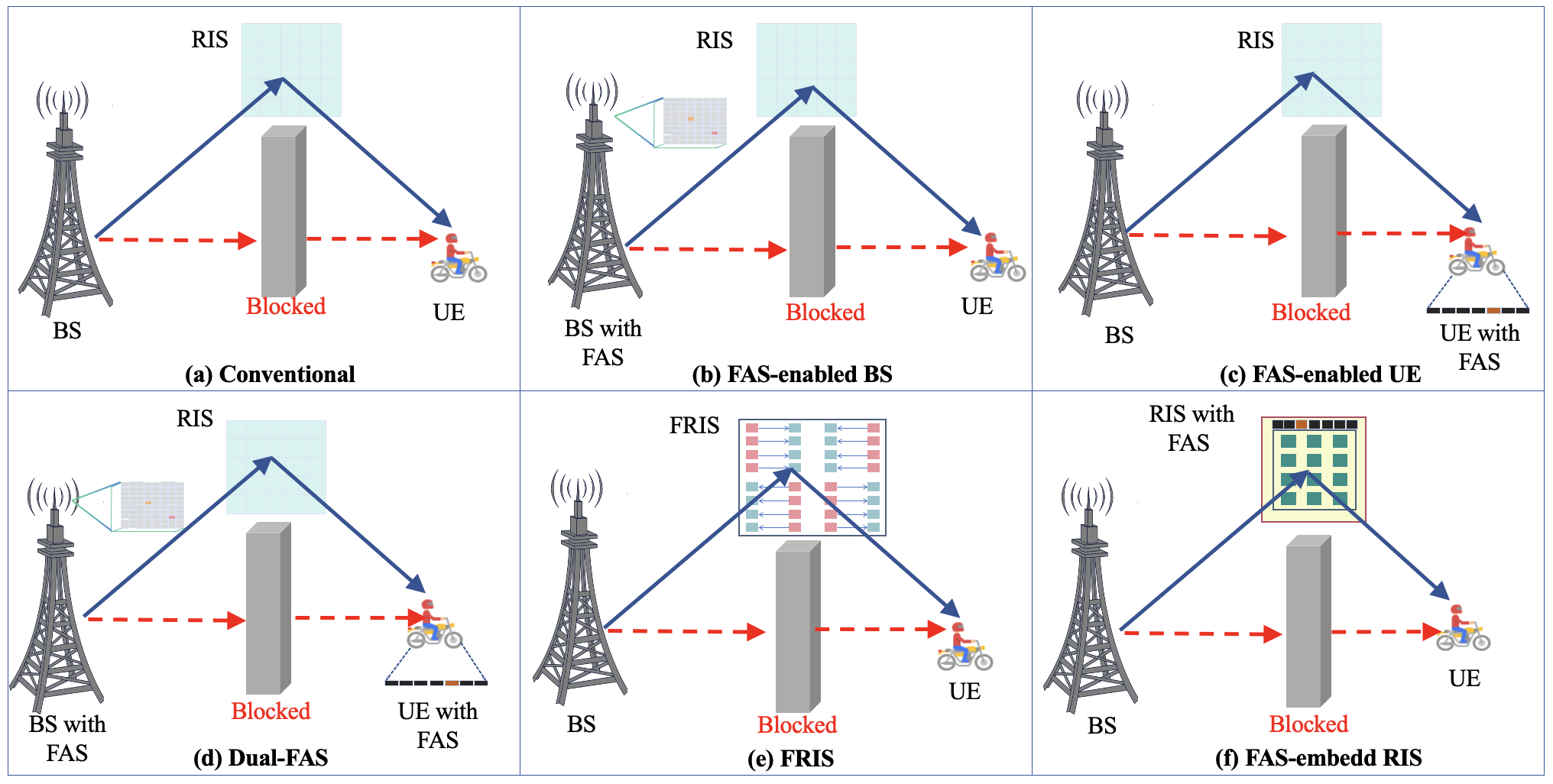}
\caption{The setups considered in the case study.}\label{fig:case-study}
\vspace{-2mm}
\end{figure*}

To validate the proposed FAS-RIS architecture and demonstrate its performance advantages, we conduct comprehensive simulations comparing six system configurations: conventional FPA-based systems (fixed BS, fixed UE, and passive RIS), FAS-enabled BS, FAS-enabled UE, dual-FAS (FAS deployed at both BS and UE), FRIS, and FAS-embedded RIS. The system comprises a BS, a RIS with $M$ reflecting elements, and a UE. The direct link between the BS and the UE is broken, necessitating the use of RIS to reestablish the communication link, as illjustrated in Fig.~\ref{fig:case-study}. Specifically, the BS, RIS, and UE are located at coordinates $(0, 0, 5)$ m, $(15, 15, 5)$ m, and $(50, 0, 0)$ m, respectively. The system operates at a frequency of $f_c = 28$ GHz with wavelength $\lambda \approx 1.07$ cm. The channels between the BS and RIS, and between the RIS and UE, are characterized by Rician fading with a path loss exponent of $2.2$ and a Rician factor of $1$. For the FAS-enabled configurations, we consider $N =10$ to $100$ ports with a spacing of $0.25\lambda$. The number of RIS elements is set to $M = 5\times 5$, while the transmit power ranges from $8$ to $20$ dBm, and the noise power is $-85$ dBm. The spatial correlation among FAS ports follows Jakes' model assuming relatively rich scattering. Monte Carlo simulations with $5000$ independent channel realizations are performed to evaluate the system performance \cite{JYao}.

For the FAS-embedded RIS configuration, we model RIS as an active relay with FAS position reconfigurability on both the receive and transmit sides, with $N_{\text{RX}} = N_{\text{TX}} = \lfloor N/4 \rfloor$ ports. The FRIS configuration optimizes element positions and spacing (from $0.4\lambda$ to $1.2\lambda$) across $\lfloor N/3 \rfloor$ configurations to maximize the channel gain. The dual-FAS system enables joint optimization of the BS and UE port selection from a reduced search space to balance performance and complexity.

\subsection{Throughput versus Number of FAS Ports}
\begin{figure}[]
\centering
\includegraphics[width=1.05\columnwidth]{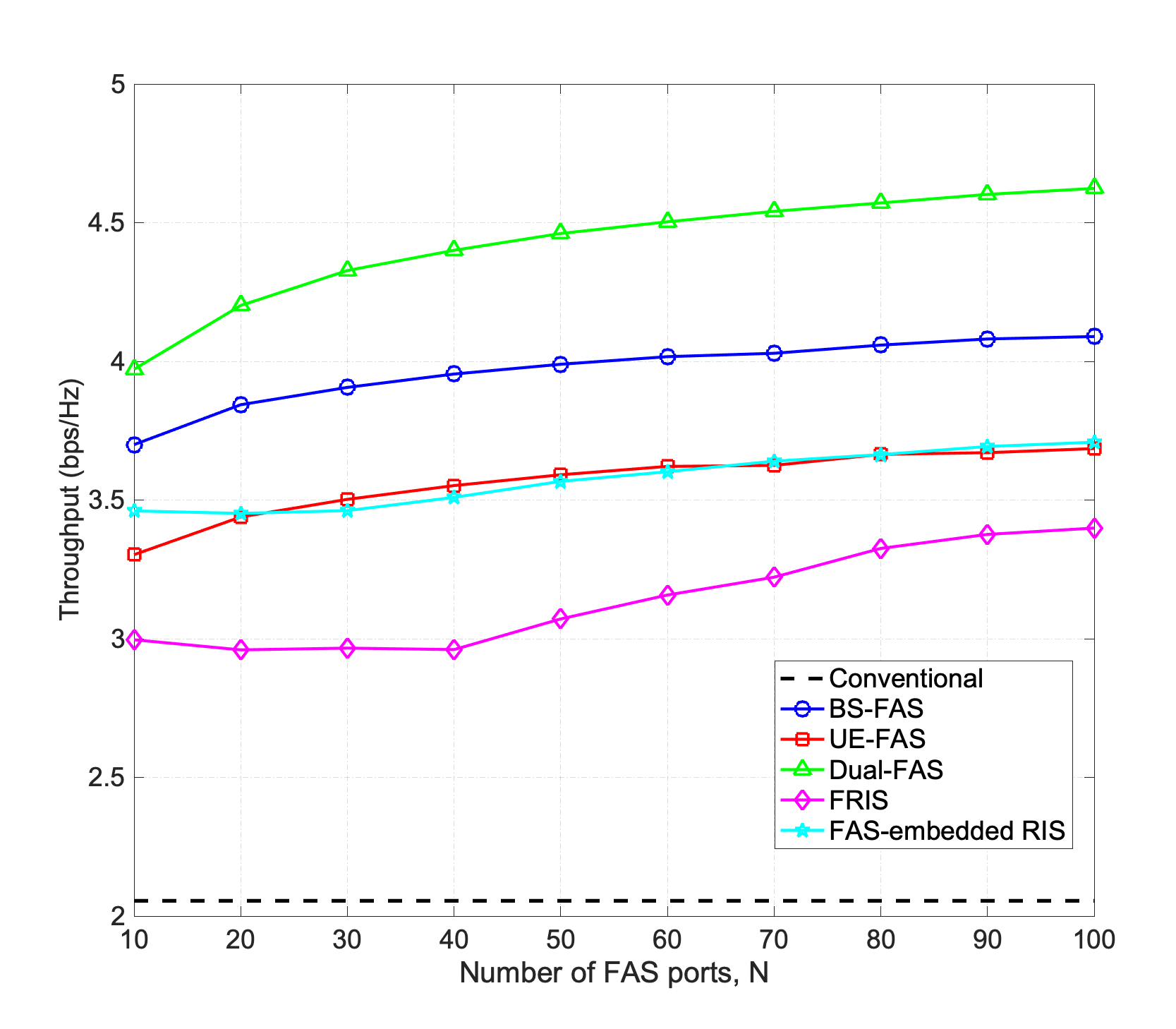}
\caption{Throughput versus the number of FAS ports $N$, with $P = 10$ dBm.}\label{fig2}
\vspace{-2mm}
\end{figure}

In Fig.~\ref{fig2}, the results are provided for the throughput performance against the number of FAS ports, $N$, assuming a fixed transmit power of $P = 10$ dBm. We observe that the dual-FAS configuration achieves the highest throughput, reaching approximately $4.6$ bps/Hz at $N = 100$ ports, demonstrating the significant benefits of enabling FAS at both the BS and UE simultaneously. This superior performance stems from the joint spatial diversity exploitation, where the system can select the optimal port combination, thereby effectively mitigating fading and maximizing the end-to-end channel gain for best performance. Additionally, the FAS-enabled BS configuration achieves approximately $4.1$ bps/Hz at $N = 100$, outperforming the FAS-enabled UE ($\sim 3.7$ bps/Hz) and FAS-embedded RIS ($\sim 3.7$ bps/Hz). This advantage arises from the BS's superior resources, including higher transmit power, larger physical aperture (approximately $0.53$ m at $N=100$), and more stable channel conditions compared to the mobile UE. The FRIS configuration, which optimizes RIS element positions and spacing, achieves $\sim 3.4$ bps/Hz, demonstrating moderate gains over the conventional system ($\sim 2.1$ bps/Hz baseline) through its reconfigurability. Notably, all FAS-enabled configurations exhibit increasing throughput with $N$, confirming the effectiveness of spatial diversity in enhancing system performance. The conventional system, lacking any reconfigurability, maintains a constant baseline performance regardless of $N$.

\subsection{Throughput versus Transmit Power}
\begin{figure}[]
\centering
\includegraphics[width=1.05\columnwidth]{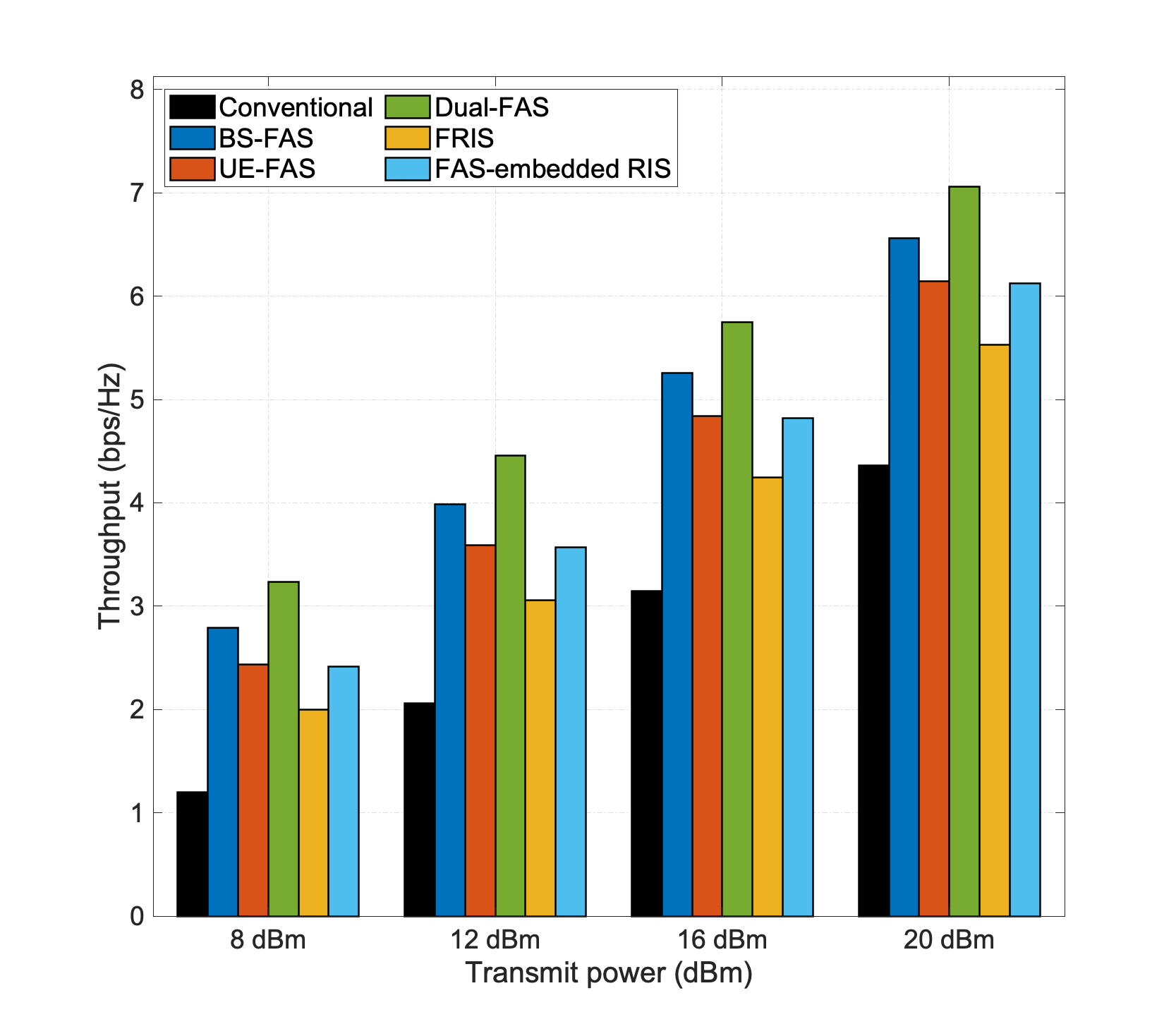}
\caption{Throughput versus transmit power $P$ with $N = 50$.}\label{fig3}
\vspace{-2mm}
\end{figure}

Fig.~\ref{fig3} illustrates the throughput performance across different transmit power levels ($8$, $12$, $16$, and $20$ dBm) with a fixed number of FAS ports $N = 50$. As expected, all configurations show monotonically increasing throughput with transmit power. At the highest power level of $20$ dBm, the dual-FAS configuration achieves approximately $7.1$ bps/Hz, followed by BS-FAS ($\sim 6.5$ bps/Hz), FAS-embedded RIS ($\sim 6.1$ bps/Hz), UE-FAS ($\sim 6.2$ bps/Hz), and FRIS ($\sim 5.5$ bps/Hz). In contrast, the conventional FPA-based system achieves only about $4.4$ bps/Hz, serving as the baseline. The performance gap between different architectures becomes more pronounced at higher power levels, highlighting the importance of FAS-enabled spatial diversity in high-SNR regimes. Specifically, the dual-FAS configuration demonstrates nearly double throughput improvement over the conventional system at $20$ dBm, while single-sided FAS configurations (BS-FAS and UE-FAS) still outperform the baseline considerably. The FAS-embedded RIS and FRIS configurations also achieve significant improvements, validating their potential as cost-effective alternatives when dual-sided FAS deployment is not feasible.

\section{Conclusions}\label{sec:conclude}
The FAS-RIS architecture represents a paradigm shift for smart-city networks, overcoming the limitations of conventional massive MIMO and purely passive RIS by tightly coupling device-level spatial agility with environment-level electromagnetic control. The proposed framework comprises five key architectural components: FAS-enabled BSs for macro-scale beamforming, FAS-enabled UE for fine-grained spatial diversity, and three RIS paradigms: FRIS, FAS-embedded RIS, and E-FAS. Combined with a two-timescale control scheme, this architecture enables transformative applications ranging from SWIPT to ISAC with ``around-the-corner'' capabilities. As we move toward 6G, the FAS-RIS framework is poised to unlock new DoFs in wireless system design, positioning it as an essential building block for intelligent, adaptive, and sensing-capable smart cities. The architectural comparison and research roadmap outlined in this article offer a clear foundation for future investigations and for the practical deployment of next-generation urban communication networks.
	

\end{document}